# Why Some Interfaces Cannot be Sharp

Naoyuki Nakagawa[1,2], Harold Y. Hwang[1,2], and David A. Muller[3]

[1]Dept. of Advanced Materials Science, University of Tokyo, Kashiwa, Chiba 277-8561, Japan

[2]Japan Science and Technology Agency, Kawaguchi 332-0012, Japan

[3]School of Applied and Engineering Physics, Cornell University, Ithaca, NY 14583, USA

**A central goal of modern materials physics and nanoscience is control of materials and their interfaces to atomic dimensions. For interfaces between polar and non-polar layers, this goal is thwarted by a polar catastrophe that forces an interfacial reconstruction. In traditional semiconductors this reconstruction is achieved by an atomic disordering and stoichiometry change at the interface, but in multivalent oxides a new option is available: if the electrons can move, the atoms don't have to. Using atomic-scale electron energy loss spectroscopy we find that there is a fundamental asymmetry between ionically and electronically compensated interfaces, both in interfacial sharpness and carrier density. This suggests a general strategy to design sharp interfaces, remove interfacial screening charges, control the band offset, and hence dramatically improving the performance of oxide devices.**

Oxide thin films have already found a variety of industrial applications ranging from mainstream electronics to niche markets such as high frequency filters. The wide variety of ground states available to the oxide family offers the potential for richer functionality than available with present conventional semiconductors – from piezoelectric resonators to magneto-optical storage. In some cases, atomic-layer control of the growth is possible, presenting new opportunities to couple different physical properties at the microscopic level. A number of recent studies have demonstrated that, when interface effects dominate, the structure and stability at small length scales introduces a host of novel considerations[1-5].

Electrostatic boundary conditions can be a dominant factor controlling the atomic and electronic structure at solid-solid interfaces. Even interfaces between formally neutral planes can have interface dipoles resulting from band offsets and bond polarizations[6,7]. However for materials with considerable ionic character, polar discontinuites introduce a larger energy cost for atomically abrupt heterointerfaces between planes of different polarity. This is the interface analog of the divergent surface energy that would result from terminating a material along a polar plane with no surface reconstruction. The consequence for growing polar materials on non-polar substrates (like GaAs on Si or Ge) is a catastrophic roughening during growth, unless the composition is graded at the interface to ensure there is no net formal interface charge[8]. This grading results in a microscopically rough interface and in many cases also a measurable electrical band offset[8,9].

How the system responds to this energy cost will have consequences for both its electrical and physical properties, such as the creation of interface phases[4,5,10] or differing interface roughness as a function of interface terminations. Much of this behavior can be captured by a simple electrostatic model which we discuss and test experimentally for (001) interfaces between $SrTiO_3$, the workhorse oxide semiconductor, and $LaAlO_3$, a closely lattice-matched insulator, useful as a gate dielectric for field-effect devices[11].

**THEORY**

The (001) planes in the $ABO_3$ perovskite structure can be divided into alternating layers of AO and $BO_2$ planes. Taking oxygen to have a formal valence of $O^{2-}$, the A and B cations can take on values of $A^{4+}B^{2+}$, $A^{3+}B^{3+}$, $A^{2+}B^{4+}$, or $A^{1+}B^{5+}$, such that the $ABO_3$ bulk structure remains neutral. Fractional charge values also arise from solid solutions and/or mixed valence states. Just as compound semiconductors made from group IV elements like Si or Ge have formally neutral (001) planes, the $A^{2+}B^{4+}O_3$ or "II-IV" structure (like $SrTiO_3$) also contains neutral AO and $BO_2$ (001) planes. An analog of the III-V or II-VI semiconductors such as GaN or CdTe that have polar planes is the $A^{3+}B^{3+}O_3$ or "III-III" structure (like $LaTiO_3$ or $LaAlO_3$), which is composed of +1 AO and -1 $BO_2$ planes.

If we consider joining perovskites from two different charge families with atomic abruptness in an (001) orientation, a polar discontinuity results at the interface. Taking the example of joining $LaAlO_3$ with $SrTiO_3$, two configurations arise, which can be defined by the composition of the layer between $AlO_2$ and $TiO_2$ at the interface – $AlO_2$/*LaO*/$TiO_2$ or $AlO_2$/*SrO*/$TiO_2$. Such a junction between polar and nonpolar planes is very common in oxide heterostructures, and the following discussion applies quite generally to many perovskite interfaces. Figures 1a and b show how an atomically abrupt interface between polar and neutral layers leads to a polar catastrophe (where the electrostatic potential diverges with thickness) if there is no redistribution of charges at the interface.

Unlike conventional semiconductors where each ion has a fixed valence, in complex oxides compositional roughening is not the only option for charge rearrangement – *mixed valence charge compensation can occur if electrons can be redistributed at lower energy cost than redistributing ions*. Conceptually, one can first construct the interface from neutral atoms, and then allow ionization, resulting in the net transfer of ½ electron per two-dimensional unit cell ($e^-$/u.c.) from $LaAlO_3$ to $SrTiO_3$ across the interface (Fig. 1c). This process leaves the overall structure neutral, with the Ti ion at the interface becoming $Ti^{3.5+}$ and the potential no longer diverges. The extra ½ electron at the $AlO_2$/*LaO*/$TiO_2$ ("*n*-type") interface should be physically detectable by transport and direct spectroscopic measurements. Indeed, metallic conductivity and Hall measurements suggest free electrons at the "*n*-type" interface[10]. Figure 1d shows the analogous construction for the $AlO_2$/*SrO*/$TiO_2$ interface where the SrO layer must now acquire an extra ½ hole per two dimensional unit cell ($e^+$/u.c.) to maintain charge neutrality – i.e. formally it should be "*p*-type". Electrically, however, this interface is insulating[10]. As this positive charge is still electrostatically necessary to avoid the divergence, and there are no available mixed valence states to compensate the ½ hole (such as $Ti^{5+}$, which is energetically inaccessible), an atomic reconstruction is required.

Here we show direct experimental evidence that the induced interface charges at the $AlO_2$/*LaO*/$TiO_2$ interfaces are compensated by mixed-valence Ti states that place extra electrons in the $SrTiO_3$ conduction band. In contrast to this *electronic* interface reconstruction, the $AlO_2$/*SrO*/$TiO_2$ interface is compensated by the introduction of oxygen vacancies at the interface, an *atomic* interface reconstruction. For repeated growth of polar interfaces in $LaAlO_3$/$SrTiO_3$ superlattices, the cations intermix rapidly to reduce the interface dipole energy, representing a fundamental limit on the growth stability of multiple polar interfaces.

**RESULTS**

To understand the electrical asymmetry between the two interface terminations we examined the interfaces with atomic-resolution electron energy loss spectroscopy (EELS) performed in a scanning transmission electron microscope (STEM) with single atom and vacancy sensitivity[12,13]. Atomic-scale measurements of composition and electronic structure at buried interfaces are possible using EELS from a high energy (200 keV) electron beam focused down to a spot as small as 1-3 Å[13-15]. By passing such a small electron beam through a thinned section, the excited states of atoms in their bulk environment can be probed. We collect only small angle inelastic scattering (<20 mrad), a regime in which the shape of the EELS signal becomes formally equivalent to that obtained from x-ray absorption spectroscopy[16]. The remaining high-angle (>50 mrad) scattering is predominantly elastic and is used to form a simultaneously-recorded annular dark field (ADF) image, with the heaviest atoms appearing the brightest[17].

Figure 2 shows the interface structure of the LaAlO$_3$ films grown on SrTiO$_3$. The structures in this study were grown by pulsed laser deposition in an ultra-high vacuum chamber on atomically flat, TiO$_2$-terminated (001) SrTiO$_3$ single crystal substrates. LaAlO$_3$, SrTiO$_3$, SrO single crystal targets, and La$_2$Ti$_2$O$_7$ polycrystalline targets were used for 2 dimensional layer-by-layer growth, as monitored by unit-cell reflection high-energy electron diffraction intensity oscillations throughout growth. We find that the AlO$_2$/*LaO*/TiO$_2$ interface is twice as rough as the AlO$_2$/*SrO*/TiO$_2$ interface (Fig. 2c). Even when the order of growth is switched and SrTiO$_3$ is grown atop LaAlO$_3$, the upper AlO$_2$/*LaO*/TiO$_2$ interface is still more diffuse than the lower AlO$_2$/*SrO*/TiO$_2$ one (Supplementary Fig. 1). Also shown in Supplementary Fig. 1 is a comparison of as-grown interfaces with those annealed in 550°C in O$_2$ for 4 hours. In all cases, we found no change in the interface atomic or electronic structure with annealing. The effect of annealing was to fill residual oxygen vacancies in SrTiO$_3$ film layers, as well as improving the crystallinity of the film, hence data from the annealed interfaces are presented here.

Figure 2d shows that as we continue to grow repeated LaAlO$_3$/SrTiO$_3$ multilayers, the interface roughness rapidly increases with each added multilayer. These structures were grown after monolayer deposition of SrO, followed by the repeated growth sequence [1 u.c. LaTiO$_3$, 4

u.c. LaAlO$_3$, 4 u.c. SrTiO$_3$]. In this manner, the superlattice is entirely composed of AlO$_2$/*LaO*/TiO$_2$ interfaces. By this construction, simple La/Sr interdiffusion cannot remove the polar discontinuity, as illustrated schematically in Fig. 2e.

The microscopic distribution of formal valences is probed using EELS where the Ti-L, O-K and La-M edges are recorded simultaneously. The Ti-L edge provides a useful fingerprint of the Ti$^{3+}$ and Ti$^{4+}$ states from which a Ti $d$ electron count and crystal field symmetry can be extracted by least squares fit to reference spectra[4,13,18,19]. The O-K edge is also sensitive to the presence of oxygen vacancies and more extended features in the band structure[13,20]. This richness makes it less amenable to an atomistic description. Instead, Fig. 3 shows that for the "*p*-type" interface the main features in the O-K edge spectra can be captured by a least squares fit to reference spectra from bulk SrTiO$_3$, bulk LaAlO$_3$, and oxygen-deficient SrTiO$_{3-\delta}$ with δ=1/4.

The lack of any statistically significant structure in the residual suggests that at our experimental sensitivity (~5-10%) and resolution (0.7 eV and 0.2 nm), the main changes in the local oxygen bonding environment at the "*p*-type" interface are due to oxygen vacancies. Excluding the oxygen-deficient reference from the fit results in a significant residual, sharply peaked at the interface. EELS line profiles like that of Fig. 3b were recorded for both interfaces. The integrated Ti, O and La counts give composition profiles, and least-squares fits to the Ti-L and O-K edge data resolve the variations in Ti valence and oxygen occupancy across the interface (Fig. 4, Supplementary Fig. 1). The quality of the fits can be evaluated from examining the lack of variation in the residuals.

The EELS-derived fractional compositions summarized in Fig. 4 resolve the puzzle over the electrical asymmetry between the "*n*-type" AlO$_2$/*LaO*/TiO$_2$ and "*p*-type" AlO$_2$/*SrO*/TiO$_2$ interfaces. We find that for the "*n*-type" interface (Fig. 4a), 0.7±0.1 excess e$^-$/u.c. are found on the Ti sites (non-zero Ti$^{3+}$) and very few oxygen vacancies (0.1±0.05 V$_O$/u.c.). There are no excess electrons (0.1±0.1) on the Ti sites for the "*p*-type" interface (Fig. 4c), where an extra ½ hole would be expected theoretically. Instead at the $AlO_2$/*SrO*/$TiO_2$ interface significant compensating oxygen vacancies (0.3±0.05 $V_O$/u.c., which would imply 0.6±0.1 fewer electrons at 2e$^-$ per O) are present (Fig. 4d) but no free holes are found (which would give a pre-peak on the O-K edge). The EELS measurements suggesting free carriers at the "*n*-type", but not the "*p*-type" interface are consistent with the electrical measurements, with the EELS deducing that the sheet carrier density at the "*n*-type" interface is close to ~5 x 10$^{14}$ cm$^{-2}$ and confined to within a few nanometers of the interface.

**DISCUSSION**

The EELS results indicate that accommodation of the polar discontinuity is the driving force for the interface electronic and atomic reconstructions we have observed. Perhaps the most striking point is that the interface with many oxygen vacancies has no excess electrons, while the interface with few oxygen vacancies has significant excess electrons. This is counter to the well known role of oxygen vacancies in bulk oxides as electron donors, indicating that the origin and the function of the interface vacancies are completely different at the interface. This is bolstered by the fact that vacancies persist despite annealing in conditions far above those necessary to fill vacancies in thick $SrTiO_{3-\delta}$ films.

In comparing the valence profiles obtained from EELS with the simple model of ½ electron/hole at the interface discussed in Fig. 1, the experimental data deviates from the model in two key respects. First, the electron are not fixed point charges, but delocalized in a screening cloud. This alters the size of the interface dipole but does not cause a divergence. Second, even though the net charge at the "*n*-type" interface is ~1/2 (0.7±0.1 e$^-$, 0.2±0.1 e$^+$) there are more electrons than expected, which in turn are compensated by slight oxygen vacancies – again altering the interface dipole.

In addition to the electrical asymmetry, why is there also an asymmetry in roughness? What we have ignored in this analysis so far is the delocalized electron cloud on the Ti sites at the $AlO_2$/*LaO*/$TiO_2$ interface. Spreading the electrons from a single plane to a few unit cells increases the interface dipole energy. This dipole can be reduced by exchanging Sr for La cations across the interface to produce a compensating dipole – i.e. physically roughening the interface. (In general an exchange of ions can only produce a dipole, but not add or remove a diverging potential). Without a delocalized screening electron or hole charge at the $AlO_2$/*SrO*/$TiO_2$ interface, there is less need for compensating cation-mixing across the interface. Furthermore, distribution of the oxygen vacancies can provide any necessary compensating dipole. Both the EELS profiles of Fig. 4 and the images of Fig. 2 show that the "*n*-type" $AlO_2$/*LaO*/$TiO_2$ interface is indeed rougher than the $AlO_2$/*SrO*/$TiO_2$ interface.

The presence of a small amount of oxygen vacancies ($\delta$=0.1±0.05) at the "*n*-type" interface (which should ideally have $\delta$=0) suggests a mechanism to reduce the band offset while still avoiding a divergence. Adding extra vacancies and compensating electrons to keep the same net charge introduces an interface dipole which will shift the band offset. Consider the case where $\delta$=0.125 and we place the missing one-in-eight O atoms in the SrO plane, i.e. the interface structure is $SrO_{0.875}$/$Ti^{3.25+}O_2$/$LaO$/$AlO_2$. This gives 0.75 excess $e^-$/u.c. at the interface, but introduces no band offset (Supplementary Fig. 2). In other words, the band offset can be tuned as a function of oxygen vacancy concentration, but with a price of adding electrons to the Ti conduction band at the interface.

A similar argument can be made for the "*p*-type" interface. The simplest interface of Fig. 1d would ideally have $\delta=0.25$ and one in four O atoms missing from the SrO plane, with no free electrons ($Ti^{4+}O_2/SrO_{0.75}/AlO_2/LaO$). The band offset can also be tuned for the "*p*-type" interface by removing O atoms and should have the same magnitude and opposite sign to the "*n*-type" case. The "*p*-type" interface with no band offset requires $\delta=0.375$ and $Ti^{3.75+}O_2/SrO_{0.625}/AlO_2/LaO$, which gives 0.25 free electrons and 3/8 missing O atoms in the SrO plane. The general case interpolates between the $\delta=0.25$ and $\delta=0.375$. Experimentally we find $\delta=0.32\pm0.06$. The absence of a $Ti^{3+}$ signal suggests the interface is closer to the ideal $\delta=0.25$ (the lower end of the oxygen error range).

Controlling the interface termination layer lets us tune between insulator and conductor, trading chemical for electrical roughness. The band offset can be further adjusted by the oxygen vacancy concentration, as discussed above, or by varying the cation ratio at the interface. Many oxide devices involve polar discontinuities at critical heterointerfaces such as in field effect devices, tunnel junctions, or ferroelectric/paraelectric interfaces. Our analysis suggests that the interface screening charges that result from the inevitable polar discontinuities are at present comparable to or larger than the largest polarizations achievable in field effect devices, and comparable to the best ferroelectric polarizations. By changing the substrate termination layer, the screening charge could be substantially reduced, which in turn should dramatically enhance the performance of these devices, possibly by orders of magnitude.

**Acknowledgements**.

We thank A. Ohtomo and M. Kawasaki for helpful discussions. This work was supported by the Mitsubishi Foundation, a Grant-in-Aid for Scientific Research on Priority Areas, and the US Office of Naval Research through the ONR EMMA MURI monitored by Colin Wood. N.N. acknowledges partial support from QPEC, Graduate School of Engineering, University of Tokyo. The Cornell Electron Microscope facilities have been supported by the NSF through the MRSEC and IMR programs.

Correspondence and requests for materials should be addressed to D.A.M. (davidm@ccmr.cornell.edu).12

**Figure Captions:**

**Fig. 1.** The polar catastrophe illustrated for atomically abrupt (001) interfaces between LaAlO$_3$ and SrTiO$_3$. **a**, The unreconstructed interface has neutral (001) planes in SrTiO$_3$, but the (001) planes in LaAlO$_3$ have alternating net charges ($\rho$). If the interface plane is AlO$_2$/*LaO*/TiO$_2$, this produces a non-negative electric field (E), leading in turn to an electric potential (V) that diverges with thickness. **b**, If the interface is instead placed at the AlO$_2$/*SrO*/TiO$_2$ plane, the potential diverges negatively. **c**, The divergence catastrophe at the AlO$_2$/*LaO*/TiO$_2$ interface can be avoided if ½ an electron is added to the last Ti layer. This produces an interface dipole that causes the electric field to oscillate about 0 and the potential remains finite. **d**, The divergence for the AlO$_2$/*SrO*/TiO$_2$ interface can also be avoided by removing ½ an electron from SrO plane in the form of oxygen vacancies.

**Fig. 2.** ADF-STEM images of the interface structures. La atoms are the brightest dots, and then Sr. The Ti atoms are the fainter dots between the Sr. The contrast is insufficient to resolve the Al atoms. **a**, LaAlO$_3$ grown on TiO$_2$-terminated SrTiO$_3$. **b**, LaAlO$_3$ grown on SrO-terminated SrTiO$_3$. **c**, Averaged line profiles across the interfaces of Fig. a,b. An error-function curve is fitted to both profiles to extract the average interface width. After accounting for a 0.2 nm probe size, the root-mean square roughness for the AlO$_2$/*SrO*/TiO$_2$ interface of Fig. 2a is $\sigma=0.77\pm0.13$ unit cells, and for the AlO$_2$/*SrO*/TiO$_2$ interface of Fig. 2b, $\sigma=1.90\pm0.11$ unit cells. **d**, A LaAlO$_3$/SrTiO$_3$ multilayer structure composed of all "*n*-type" interfaces shows a progressive increase in interface roughness with growth. **e**, A schematic of the cation intermixing observed in Fig. 2d.

**Fig. 3.** O-K edge EELS profile across a AlO$_2$/*SrO*/TiO$_2$ "p-type" interface. **a**, The ADF image of the interface – SrTiO$_3$ is the darker material in the upper half of the image, and LaAlO$_3$ film is in the lower half (1 nm scale bar). **b**, O-K EELS spectra recorded from the circles in Fig. 3a – gray curves. Bulk reference bulk spectra of SrTiO$_3$ (red), LaAlO$_3$ (blue) and $\delta=1/4$ SrTiO$_{3-\delta}$ (green) are shown at the bottom of the panel. The colored lines show least-squares fits to the position-dependent spectra color-coded by the fractional contribution of the red, green and blue reference spectra. **c**, The La-M edge (simultaneously recorded with the O-K data) showing the interface is graded over two unit cells. **d**, The bottom panel shows in detail the decomposition of the O-K edge at the interface. Experimental data are shown in black dots, and the violet curve is the fit from the addition of the three reference spectra. The residual to the fit is given by the black line at the bottom.

**Fig. 4.** Chemical profiles of LaAlO$_3$ on (001) SrTiO$_3$ for both interface terminations. **a**, AlO$_2$/*LaO*/TiO$_2$ interface showing the fractions of elemental Ti and La from the Ti-L and La-M edges, as well as the Ti$^{3+}$ fraction determined from a least squares fit to the Ti-L edge from Ti$^{3+}$ and Ti$^{4+}$ reference spectra. There is excess Ti$^{3+}$ on the substrate side of the interface. **b**, Corresponding Ti and La EELS profiles for the AlO$_2$/*SrO*/TiO$_2$ interface, showing almost no excess Ti$^{3+}$. **c**, Fractional compositions from the least squares fit to the O-K edge profile for the AlO$_2$/*LaO*/TiO$_2$ terminated interface, showing a net vacancy excess of $\delta=0.1\pm0.04$. **d**, The O-K edge fractional composition (from Fig. 3b) for the AlO$_2$/*SrO*/TiO$_2$ interface showing a significant accumulation of excess vacancies ($\delta=0.32\pm0.06$). The root mean square residual to the fits is shown below each plot.

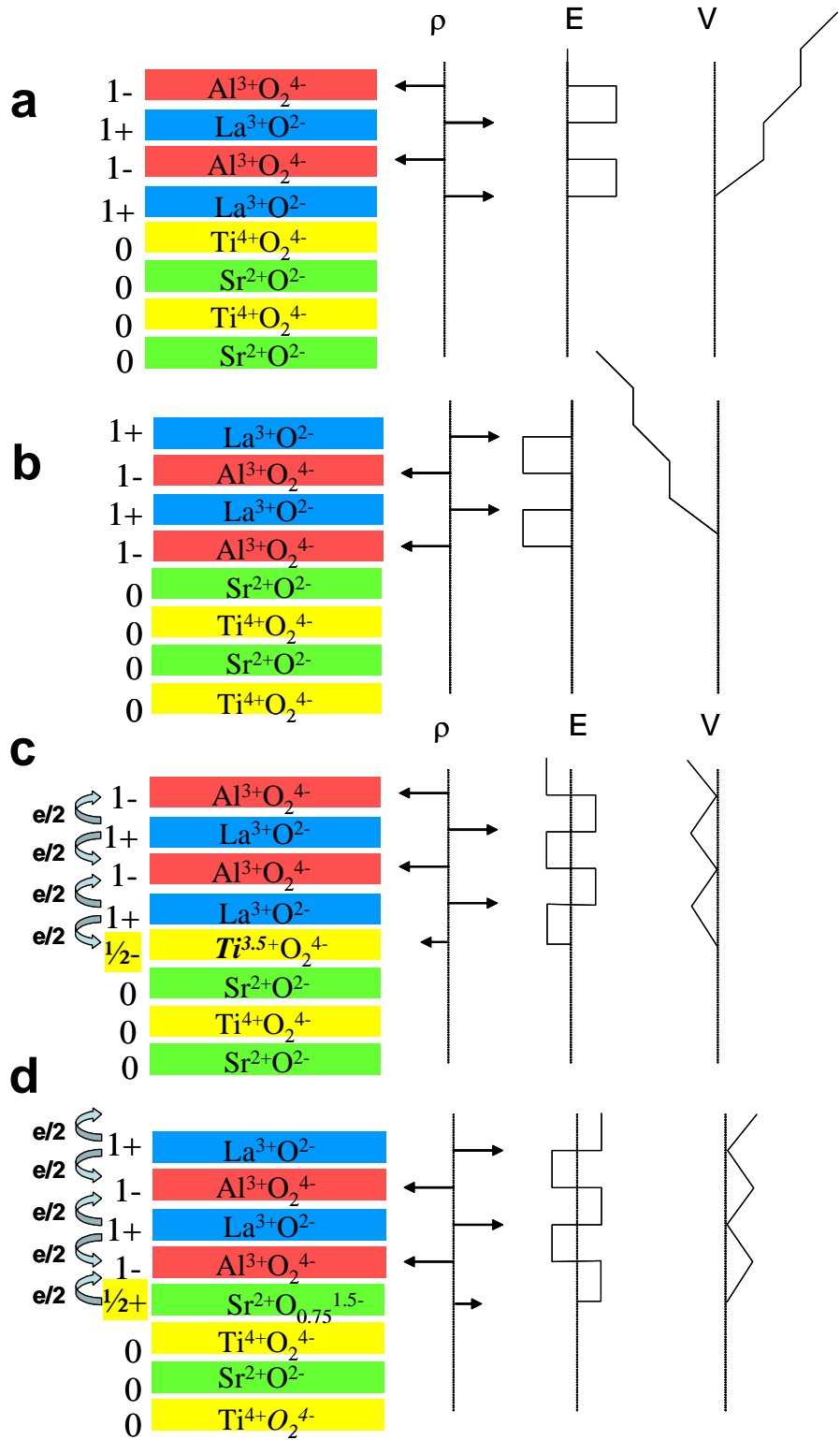

Figure 1. Nakagawa, Hwang and Muller

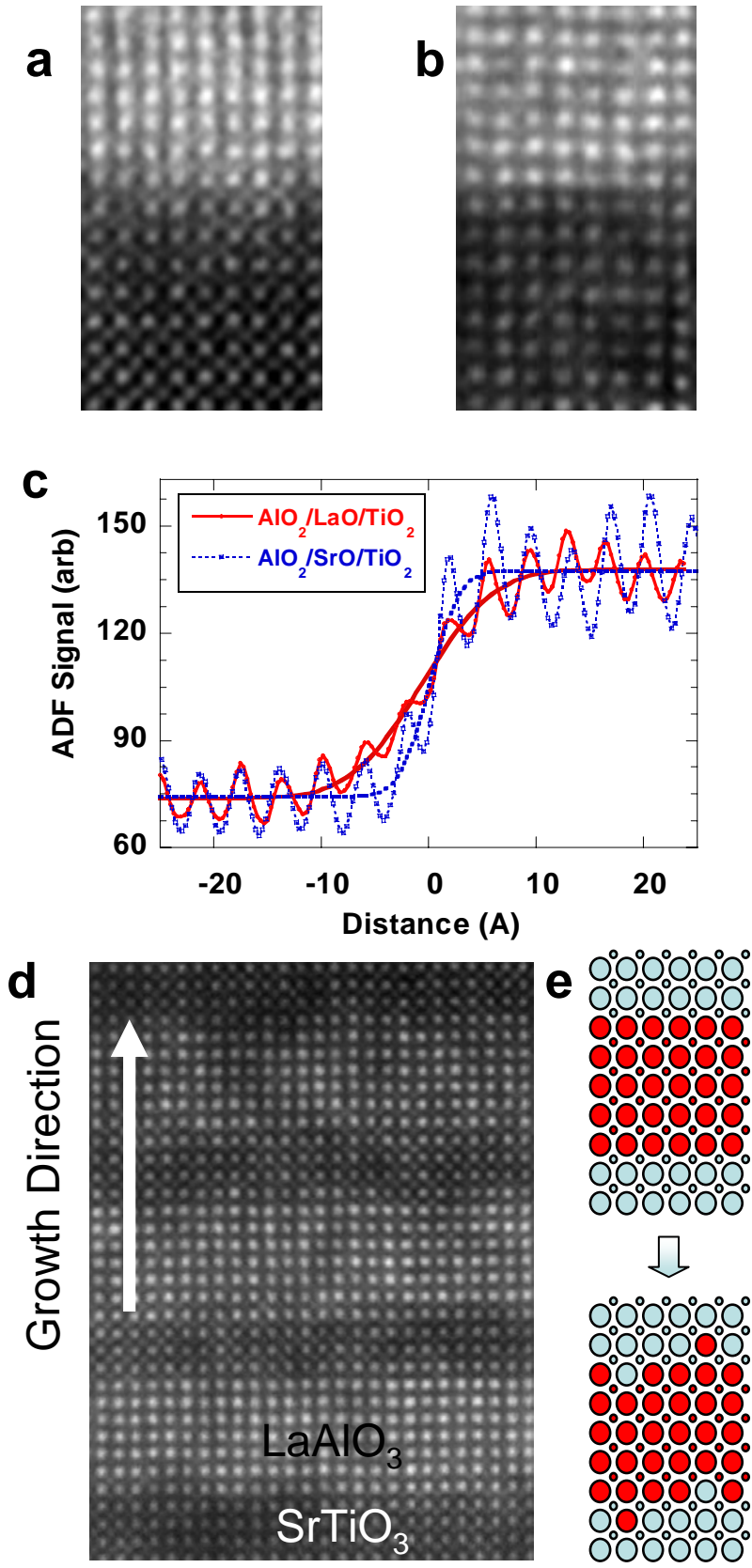

Figure 2. Nakagawa, Hwang and Muller

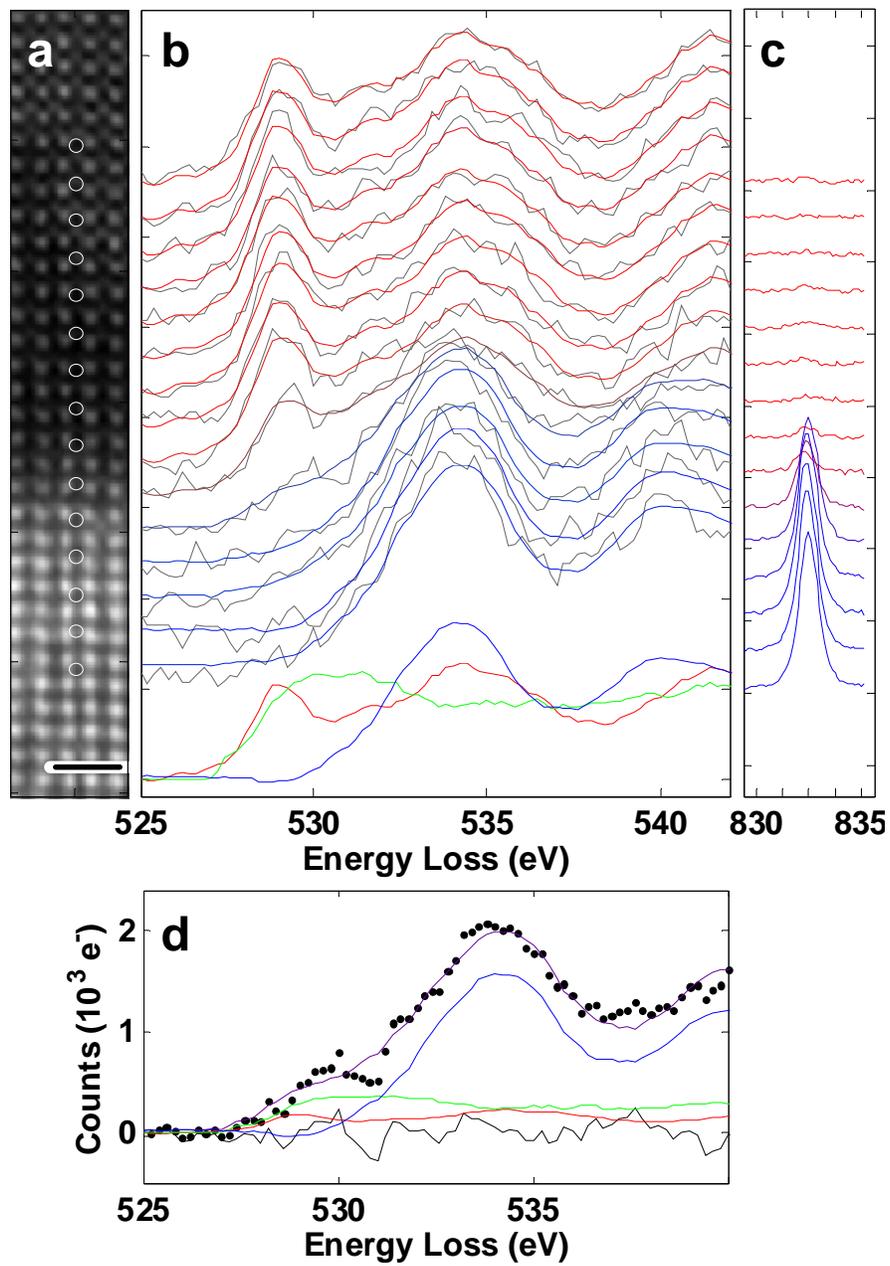

Figure 3. Nakagawa, Hwang and Muller

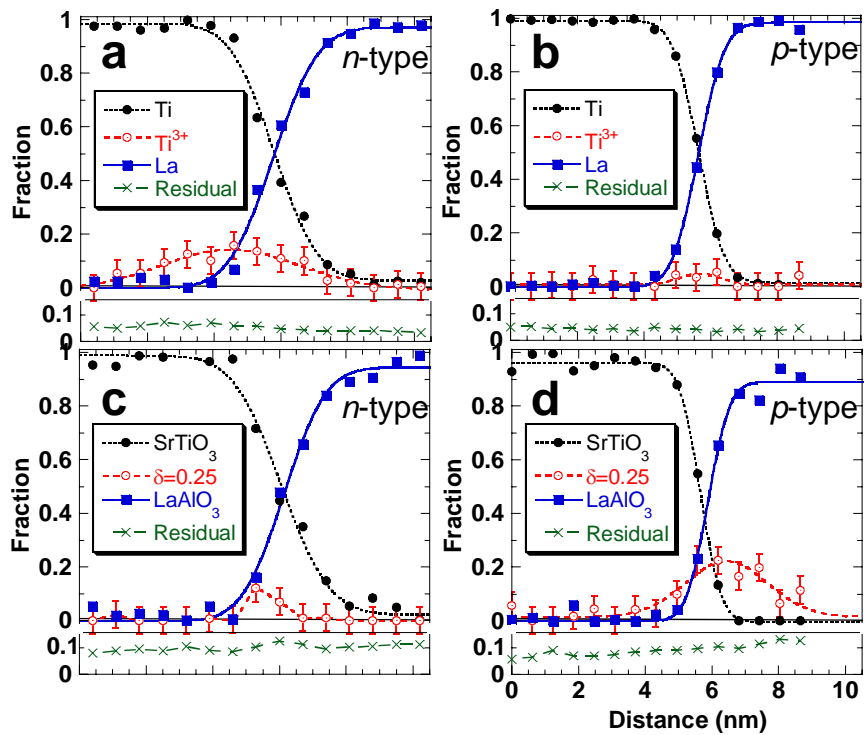

Figure 4. Nakagawa, Hwang and Muller

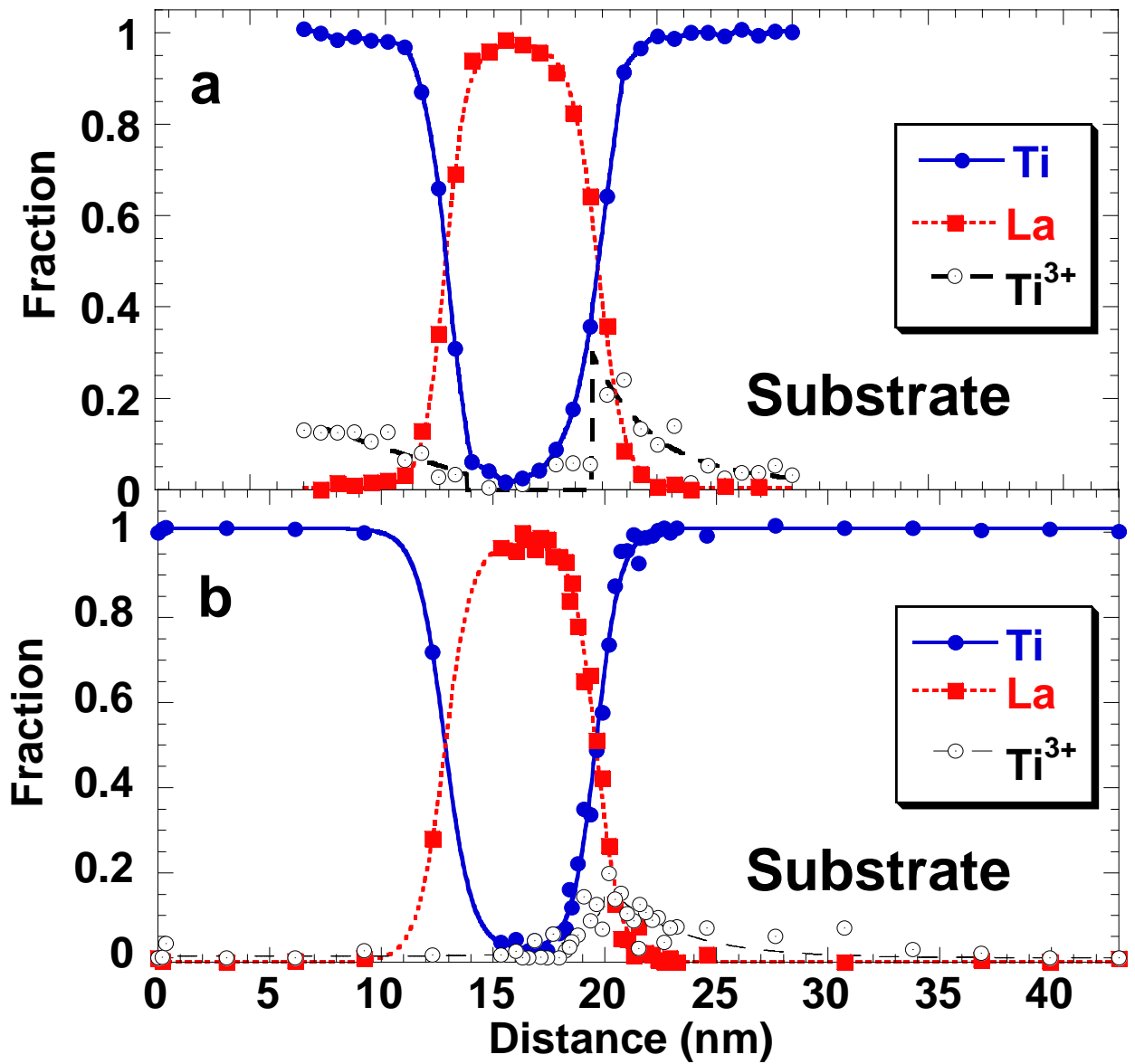

**Supplementary Figure 1.** Nakagawa, Hwang and Muller

SrTiO$_3$/LaAlO$_3$/SrTiO$_3$ multilayer before **a**, and after **b**, annealing. The growth direction is right to left. The lower interface remains metallic and unchanged after annealing. The upper interface was insulating (no Ti$^{3+}$) before annealing, even though the SrTiO$_3$ layer above it contained oxygen vacancies. After annealing, the vacancies have been filled. Also note from (a) that the upper interface (*p*-type) is less diffuse than the lower (*n*-type).

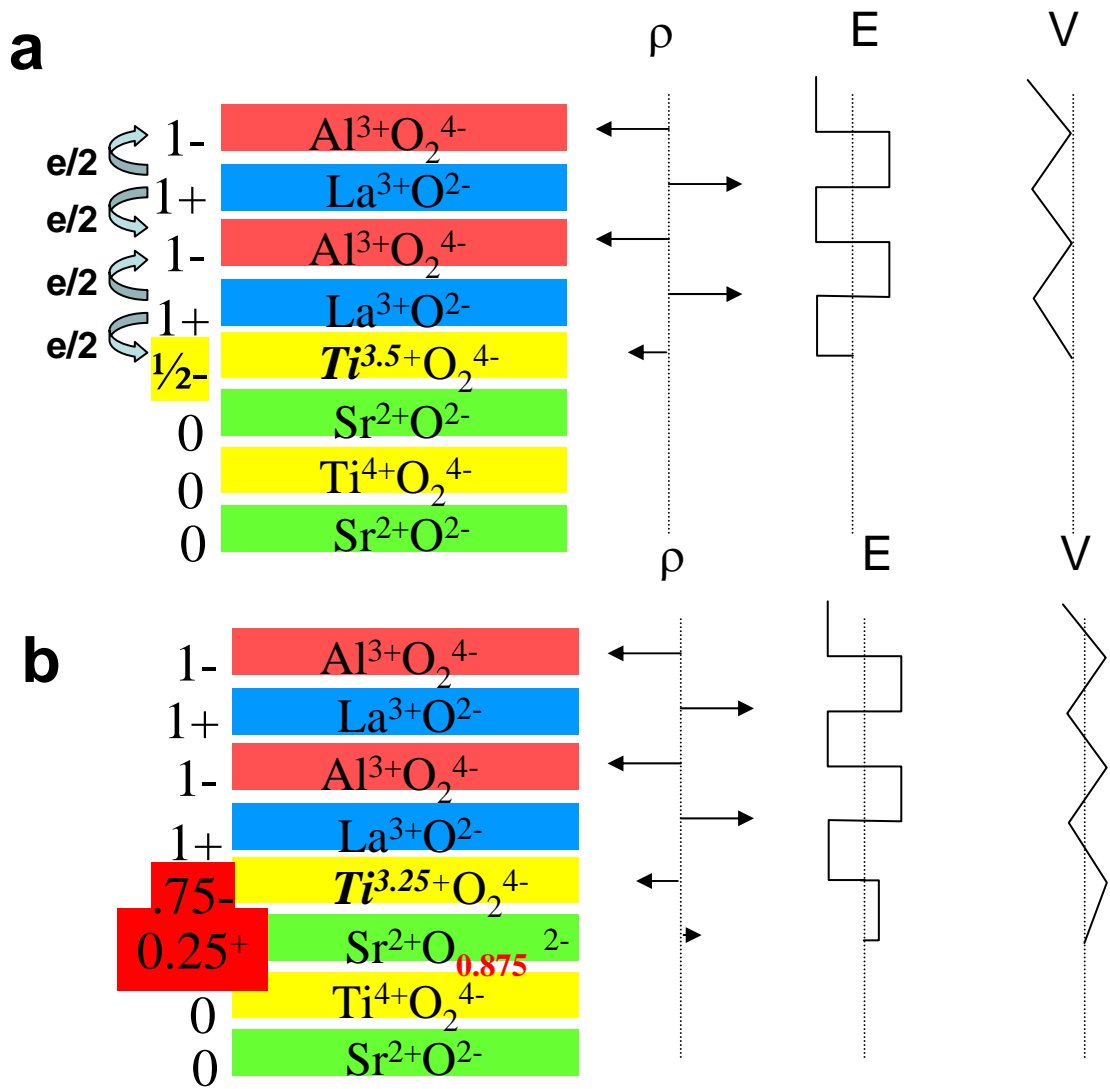

**Supplementary Figure 2.** Nakagawa, Hwang and Muller

Tuning the band offset. **a**, shows the *n*-type interface of Fig. 1c where the average potential on the LaAlO$_3$ side is shifted by 0.46 V. **b**, The band offset between SrTiO$_3$ and LaAlO$_3$ can be removed by adding an interface dipole composed of an extra 0.25 e$^-$ on the interfacial TiO$_2$ layer, and an extra 0.125 oxygen vacancies on the next SrO layer. Overall the net interface charge must remain at 0.5 to avoid a potential divergence.